\documentclass[preprint,amsmath,amssymb,aps,prd,showpacs]{revtex4-1}
 \pdfoutput=1   

\usepackage{float}

\usepackage{epsfig}
\usepackage{dcolumn}
\usepackage{bm}
\usepackage{tikz}
\usepackage{graphicx, graphics, color}
\usepackage{dcolumn}
\usepackage{bm}
\usepackage{hyperref}
\usepackage[mathlines]{lineno}
\usepackage{float}
\usepackage{subfig}
\usepackage{tabularx}

\newcommand{\be}{\begin{equation}}
\newcommand{\ee}{\end{equation}}
\newcommand{\ben}{\begin{eqnarray}}
\newcommand{\een}{\end{eqnarray}}

\graphicspath{{figs/}}
\graphicspath{{../figs/}}

\begin{document} 

\title{Multiple mountains on a pulsar: implications for gravitational waves and the spin-down rate}

\author{Paritosh Verma} 
\affiliation{Institute of Physics, 
Maria Curie-Sklodowska University, 
pl.~Marii Curie-Sklodowskiej 1,  20-031 Lublin,  Poland}
\email{pverma@kft.umcs.lublin.pl}

\author{Sudip Bhattacharyya} 
\email{sudip@tifr.res.in}
\affiliation{Department of Astronomy and Astrophysics, \\Tata Institute of Fundamental Research, \\
1 Homi Bhabha Road, Colaba, Mumbai 400005, India}%

\date{\today}

\begin{abstract}
A pulsar, i.e., a spinning neutron star, with a deformation could emit gravitational waves continuously.
Such continuous waves, which have not been detected yet, will be very useful to study gravitational physics and to probe the extreme physics of neutron stars.
While typically such waves from a pulsar are estimated considering an overall stellar ellipticity, there can be multiple irregularities or mountains in the stellar crust that the gravity of the star cannot smooth. 
In this paper, we consider this realistic situation and compute the strain, power, torque and the pulsar spin-down rate due to multiple mountains supported by the stellar crust.
Here, we consider astronomically motivated mountain distributions and 
use the Brans-Dicke theory of gravity which has three polarization states: two tensors dominated by the time-varying quadrupole moment and one scalar dominated by the time-varying dipole moment.
We also give the limiting results for general relativity.
\end{abstract}

\maketitle
\flushbottom

\section{Introduction}
\label{sec:intro}


The LIGO and Virgo detectors have already observed gravitational radiation from the merger of compact binary objects, but there is still a quest to search for other sources of such radiation. Another source of gravitational waves (GW) is a tiny deformation in the crust of a spinning neutron star (NS), which may manifest as a  pulsar. This deformation or mountain can be as high as a few cm, induces asymmetry in the star about the spin axis, and produces a time-varying mass quadrupole moment. This deviation from the symmetry is responsible for almost a pure sinusoid GW signal with the frequency proportional to the star's spin frequency in the source frame. This signal from an isolated NS is much weaker than that originating from the binary merger, but its long-lasting nature may lead to detection with future detectors \cite{Hask-Bej, Piccini}.

The model of pulsar-emitting continuous GWs in general relativity (GR) has been studied in detail \cite{JKS1, JK1, JK2, Jones2010}, and various search methods incorporated into LIGO-Virgo-Kagra (LVK) pipeline to look for these signals. Some important search methods to find these signals are $\mathcal{F}/ \mathcal{G}$-statistic \cite{JKS1}, 5n-vector method \cite{astone2012} and Bayesian analysis \cite{pitkin2017}. 

Although GR has passed several tests, it has some flaws which has made the scientific community think about alternative theories of gravity as well as testing GR \cite{GR test GWTC3, GR-test-GW170817, Wex, Will2014}. For instance, quantization of GR  and the nature of dark energy  are still an open issue. In this paper, we implement the Brans–Dicke (BD) theory \cite{Brans-Dicke, Jordan, Fierz,  Misner, Fuji-Maeda}, which comes under the class of scalar-tensor theories. This theory still has general coordinate invariance besides having an additional degree of freedom. This additional degree of freedom is the scalar field $(\phi(x))$ due to which the gravitational ``constant" G is not a constant anymore but rather depends on position and time, as proposed by Paul Dirac. There have been attempts to use this scalar field to replace occult fluids like dark matter and dark energy \cite{Hongsu, Bertolami, JBD-cosmo, Sola}. The field equations of the BD theory incorporate a parameter called the BD coupling constant $\omega_{\rm BD}$ which could be estimated through experiments. 

In the past, there have been attempts to study gravitational waves in scalar-tensor theories of gravity. Radiation emitted in scalar and tensor waves from a binary system was studied in \cite{Will-1977}. Moreover, recently, there has been an attempt to study continuous GWs from pulsars in BD theory \cite{Verma}. The author considered a model in which a mountain is present on the equator of a pulsar and calculated the polarizations. Furthermore, a new statistic called the $\mathcal{D}$-statistic was developed to search for GW signals. $2 \times \mathcal{D}$-statistic is a $\chi^2$ distribution with 2 degrees of freedom, and it generalizes the well-known $\mathcal{F}$-statistic mentioned previously. This $\mathcal{D}$-statistic was tested by Monte Carlo simulation and then finally implemented in the LVK pipeline to search for scalar waves \cite{BD-LVK} \cite{BD-LVK1}. 

In this paper, we extend the model discussed in \cite{Verma} to calculate strains for GW polarizations and the emitted power when multiple mountains are present on the pulsar's surface. 
For simplicity, we assume the pulsar to be perfectly spherical with some tiny deformations on its surface. 
The perfect sphere assumption is reasonable because even for rapidly spinning neutron stars, 
the polar radius is only a few percent ($< 4$\%) less than the equatorial radius in most cases \cite{SB-2017}, \cite{NICER-Riley}.
Besides, note that the multiple mountains scenario could be more realistic than a single mountain on the neutron star because if one mountain can form due to irregularities which cannot be smoothed by stellar gravity, then many mountains could also form.
Furthermore, while the crust could support a relatively high deformation (e.g., a net ellipticity of $\epsilon \sim 10^{-7}$ or higher; e.g., \cite{BH-2006}), the $\epsilon$ of millisecond pulsars, i.e., rapidly spinning neutron stars, could be $\sim 10^{-9}$ for many sources (e.g., \cite{BH-2017}, \cite{Woan-2018}, \cite{SB-2020}).
Such low $\epsilon$ values could be explained if multiple mountains spread over the neutron star, since in such cases their effects may somewhat balance each other, and the net $\epsilon$ value could be smaller.

We discuss the formulae of gravitational wave strains in section~\ref{sec:pol}, and present the expressions of radiated power, spin-down rates, and other parameters in  section~\ref{sec:power}.
Results using general relativity are given in section~\ref{sec:GR}.
We discuss the special case of the single-harmonic model and spin-down limit in section~\ref{Single_Harmonic}.
In sections~\ref{mountain} and \ref{numerical}, we consider various distributions of mountains on the pulsar, and in section~\ref{conclusion}, we give concluding remarks.


\section{\label{sec:pol} Gravitational Wave Strains for different Polarizations}

In order to demonstrate the effects of multiple mountains in a simple way, 
we assume that the mountains on the pulsar are distributed in a regular manner.
Let the mass of the $k^{th}$ mountain be $m_k$ and its coordinates are

\begin{eqnarray} 
\label{eq:1} \nonumber
x_k &=&   a \sin \theta_k \cos \phi_k, \\ \nonumber
y_k &=& a \sin \theta_k \sin \phi_k, \\ 
z_k &=& a \cos \theta_k,
\end{eqnarray} 
where, $a$ is the radius of the star, $ \theta_k \in [0, \pi]$ and $ \phi_k \in [0, 2 \pi]$.

The BD theory comprises three transverse polarization states, out of which two are tensor modes and one is the scalar mode. In the language of particle theory, the tensor modes can be thought of as spin-2 particles, whereas the scalar mode can be thought of as spin-0 particles \cite{Poisson, Isi, Verma-rod-BD}.

The GW strains for different polarizations in BD theory are given by \cite{Verma}

\label{eq:2}
\begin{equation}
h_{+} (t) = \frac{G}{r c^{4}} (1 - \zeta) 
(\ddot{Q}^{xx}_{W} (t^{\prime}) - \ddot{Q}^{yy}_{W} (t^{\prime})), 
\end{equation}

\begin{equation}
h_{\times} (t) = \frac{2 G}{r c^{4}} (1 - \zeta)  \ddot{Q}^{xy}_{W} (t^{\prime}), \label{eq:2.1}
\end{equation}

and
\begin{equation}
h_{S} (t) = \frac{2G}{r c^{2}}  \zeta \left[  M(t^{\prime})  +  \frac{1}{c} \dot{D}_{W}^{z} (t^{\prime})  - \frac{1}{2 c^2}  \ddot{Q}^{zz}_{W}  (t^{\prime}) \right],
\label{eq:2.2}
\end{equation}
where, $h_{+} (t)$, $h_{\times} (t)$, and $h_{S} (t)$ are for the plus, cross, and scalar polarizations, respectively. 
Here, $r$ is the distance of the source from the detector, $t$ is the time when the polarizations are measured, $t'$ represents the retarded time, $c$ is the speed of light in vacuum, $G$ is the gravitational constant, 
${Q}^{ij}_{W} (t^\prime)$ is the moment of inertia tensor of the source in the wave frame, $D_W(t^\prime)$ is the dipole moment of the source in the wave frame, and $M(t^\prime)$ is the mass monopole contribution. 
To simplify the calculations, $\zeta$ is defined as

\begin{equation}
\zeta \equiv \frac{1}{2 \omega_{\rm BD} + 4}.
\label{eq:2.3}
\end{equation}
 Here, $\omega_{\rm BD}$ is a free parameter of the field equations of the BD theory. 
 The Cassini mission suggests that massless scalar-tensor theories must have $\omega_{\rm BD} > 40,000$
 \cite{Cassini}.
 Using this lower limit and Eq. (\ref{eq:2.3}), one obtains $ \zeta<  0.0000125$.
 Note that
$\zeta \rightarrow 0 $ implies the GR regime with only two tensor polarizations and no scalar polarization.

The mass monopole radiation could be produced by astrophysical sources with varying
gravitational mass \cite{gw-monopole}. 
 Since we are considering an isolated pulsar with a non-varying mass, we shall drop the mass monopole term. First, we calculate the dipole and quadrupole moments of the system in the source frame and then move to the wave frame using the orthogonal rotation. The symmetric trace-free quadrupole moment tensor in the source frame is given by

 \begin{equation}\label{eq:3}
Q_{s}^{ij} = \int \rho_k \left[ x^i x^j - \frac{1}{3} r^2 \delta_{ij}    \right] dx dy dz,
 \end{equation} 
where, $r^2 = x^2 + y^2 + z^2$ and $\rho_k$ is the density of the $k^{th}$ mountain. The size of a mountain is much smaller than the dimensions of the star and we can approximate it to $m_k = M \epsilon_k$ where M is the mass of the pulsar and $\epsilon_k$ is the ellipticity of $k^{th}$ mountain. Hence, we can safely approximate a mountain by the Dirac Delta function and the density of the $k^{th}$ mountain is given by

 \begin{equation}\label{eq:4}
\rho_k = \epsilon_k M \delta (x-x_k) \delta (y-y_k) \delta (z-z_k)
 \end{equation} 
This, after some algebraic calculations, yields:
\begin{widetext}
\begin{equation}
Q_{s}^{ij}  =   M a^2   
 \begin{bmatrix} \sum_{k=1}^{N} \epsilon_k (\sin^2 \theta_k \cos^2 \phi_k -\frac{1}{3} )  & \sum_{k=1}^{N} \epsilon_k \frac{1}{2} \sin^2 \theta_k \sin 2 \phi_k  & \sum_{k=1}^{N} \epsilon_k \frac{1}{2} \sin 2 \theta_k \cos \phi_k   \\ \sum_{k=1}^{N} \epsilon_k \frac{1}{2} \sin^2 \theta_k \sin 2 \phi_k & \sum_{k=1}^{N} \epsilon_k (\sin^2 \theta_k \sin^2 \phi_k -\frac{1}{3} ) & \sum_{k=1}^{N} \epsilon_k \frac{1}{2} \sin 2 \theta_k \sin \phi_k \\ \sum_{k=1}^{N} \epsilon_k \frac{1}{2} \sin 2 \theta_k \cos \phi_k & \sum_{k=1}^{N} \epsilon_k \frac{1}{2} \sin 2 \theta_k \sin \phi_k & \sum_{k=1}^{N} \epsilon_k ( \cos^2 \theta_k  -\frac{1}{3} )  \end{bmatrix}
 \label{eq:5}
\end{equation}
\end{widetext}
where, $N$ is the total number of mountains. Appendix \ref{appdx} presents a detailed calculation to obtain the components of the moment of inertia tensor. 

Similarly, the dipole moment in the source can be calculated using

    \begin{equation} 
 \label{eq:6}
D_s^i  =  \int \rho_k x^i dx dy dz,
 \end{equation} 
and the result is
\begin{equation} 
 \label{eq:7}
D_s  =  Ma \begin{bmatrix} \sum_{k=1}^{N} \epsilon_k \sin \theta_k \cos \phi_k   \\ \sum_{k=1}^{N} \epsilon_k  \sin \theta_k \sin \phi_k   \\ \sum_{k=1}^{N} \epsilon_k \cos \theta_k  \end{bmatrix}.
 \end{equation} 

To calculate the GW strains for different polarizations, we need to move to the wave frame. 
Following the construction in chapter 2.5 of \cite{Krolak},  we have
\begin{equation} \label{eq:7} 
   D_{W}(t) = S \cdot R(t) \cdot D_{s}
\end{equation}
and
\begin{equation} \label{eq:8} 
   Q_{W}(t) = S \cdot R(t) \cdot Q_{s} \cdot R(t)^{T}\cdot S^{T},
\end{equation}
where $S$ is the transformation matrix from the source frame to an inertial frame, and $R(t)$ is the transformation matrix from the inertial frame to the wave frame.
$S^T$ and $R(t)^T$ represent the transposes of $S$ and $R$, respectively. 
The matrix $R(t)$ is given by
\begin{equation} \label{eq:9}
   R(t) =   \begin{bmatrix} \cos \omega t & -\sin \omega t & 0 \\ \sin \omega t & \cos \omega t  & 0 \\ 0 & 0 &1 \\ \end{bmatrix},  
\end{equation}
where $\omega$ is the angular velocity of the star.
The matrix $S$ has the form
\begin{equation} \label{eq:10} 
   S  =    \begin{bmatrix} \cos \iota & 0 & -\sin \iota   \\ 0 & 1  & 0 \\ \sin \iota & 0 & \cos \iota \\ \end{bmatrix},
\end{equation}
where $\iota$ is the angle between the angular momentum vector of the rotating neutron star
and the direction along which the wave travels.

After some algebraic manipulations and ignoring the mass monopole term in the scalar wave, we obtain the following result.

\begin{equation} \label{eq:11}
\begin{split}
h_+(t) & = - \frac{2G (1-\zeta)}{rc^4} \omega^2 M a^2  \left( 1 + \cos^2 \iota \right) \sum_{k=1}^{N} \epsilon_k  \cos \left( 2 \omega t^\prime + 2 \phi_k  \right)   \sin^2 \theta_k   \\
 & + \frac{G (1-\zeta)}{2rc^4} \omega^2 M a^2 \sin 2 \iota  \sum_{k=1}^{N} \epsilon_k  \cos \left(  \omega t^\prime +  \phi_k  \right) \sin 2 \theta_k,  
\end{split}
\end{equation}

\begin{equation} \label{eq:12}
\begin{split}
h_{\times}(t) & = -\frac{4G (1-\zeta)}{rc^4} \omega^2  M a^2 \cos \iota  \sum_{k=1}^{N} \epsilon_k \sin \left( 2 \omega t^\prime + 2 \phi_k  \right) \sin^2 \theta_k   \\
 & + \frac{G (1-\zeta)}{rc^4} \omega^2 M a^2 \sin  \iota \sum_{k=1}^{N} \epsilon_k \sin \left(  \omega t^\prime +  \phi_k  \right)   \sin 2 \theta_k,
\end{split}
\end{equation}

and
\begin{equation} \label{eq:13}
h_{S} (t) \approx - \frac{2G}{rc^3} \zeta \omega M a \sin \iota \sum_{k=1}^{N} \epsilon_k \sin \theta_k \sin \left( \omega t^\prime + \phi_k \right).
\end{equation}

It should be noted that tensor waves are emitted at two frequencies: the spin-frequency of the pulsar and twice the spin-frequency of the pulsar. 
The first term in both $h_+(t)$ and $h_{\times}(t)$ corresponds to twice the spin-frequency, whereas the second term corresponds to the spin-frequency. However, scalar waves are emitted only at the spin frequency.

\section{\label{sec:power} Radiated Power}

The total power (${P_{\rm grav}}$) emitted in gravitational radiation per unit area is given by \cite{Will-1977,Verma-rod-BD}

\begin{equation}
\label{eq:14}
\frac{dP_{\rm grav}}{dA} = \frac{c^3}{16 \pi G (1 - \zeta) } <   \dot{h}^2_{+}(t) + \dot{h}^2_{\times}(t) +  \left( \frac{1 - \zeta}{\zeta} \right)  \dot{h}_{s}^2(t)   > 
\equiv \frac{dP^{(T)}}{dA} + \frac{dP^{(S)}}{dA},
\end{equation} 
where, $dA = r^2 \int _{\rho = 0}^{2 \pi} \int _{\iota = 0}^{ \pi}  \sin \iota d \iota d \rho  $ is the area element in spherical coordinates and $ < \cdot > $ implies the time average.
Here, $\frac{dP^{(T)}}{dA} $ is the power emitted in the tensor wave 

\begin{equation} \label{eq:15}
 \frac{dP^{(T)}}{dA}  \equiv \frac{c^3}{16 \pi G(1 - \zeta)} <   \dot{h}^2_{+}(t) + \dot{h}^2_{\times}(t) >,  
\end{equation} 
and $\frac{dP^{(S)}}{dA} $ is the power emitted in the scalar wave.  

 \begin{equation} \label{eq:16}
 \frac{dP^{(S)}}{dA} \equiv \frac{c^3}{16 \pi \zeta G  } <   \dot{h}^2_{s}(t) >
\end{equation} 
On substitution of Eqs. (\ref{eq:11}), (\ref{eq:12}) and (\ref{eq:13}) into Eqs. (\ref{eq:15}) and (\ref{eq:16}), we obtain the following expressions of power emitted in tensor and scalar polarizations in the presence of multiple mountains:

\begin{widetext}
\begin{eqnarray}
P^{(T)} = && \frac{32}{5} \frac{G(1-\zeta)}{c^5} M^2 a^4  \omega^6  \left[ \sum_{k=1}^{N} \sum_{j=1}^{N} \epsilon_k \epsilon_j \sin^2 \theta_k \sin^2 \theta_j \cos(2 \phi_k - 2 \phi_j) \right]  \nonumber \\
 && + \frac{1}{10} \frac{G(1-\zeta)}{c^5} M^2 a^4 \omega^6  \left[  \sum_{k=1}^{N} \sum_{j=1}^{N} \epsilon_k \epsilon_j \sin 2 \theta_k \sin 2 \theta_j \cos( \phi_k -  \phi_j) \right]
 \label{eq:17.1}
\end{eqnarray}
\end{widetext}

and
 \begin{equation} \label{eq:18.1}
 P^{(S)} = \frac{1}{3} \frac{G \zeta}{c^3}  M^2 a^2 \omega^4 \sum_{k=1}^{N} \sum_{j=1}^{N} \epsilon_k \epsilon_j \sin \theta_k \sin \theta_j \cos (\phi_k - \phi_j).
\end{equation} 
Here, we have used $<\sin^2 nt> = <\cos^2 nt> \approx \frac{1}{2}$ and $< \sin nt \cos mt> \approx 0$. Also, $< \sin nt \sin mt>  = < \cos nt \cos mt> \approx 0 \  \text{when} \  m \neq n$. 
These relations hold for any integral values of m and n. 
In order to compactify the expressions, we define net ellipticities as

\begin{equation}\label{eq:18.2}
\epsilon_{2q} \equiv \left[ \sum_{k=1}^{N} \sum_{j=1}^{N} \epsilon_k \epsilon_j \sin^2 \theta_k \sin^2 \theta_j \cos(2 \phi_k - 2 \phi_j) \right]^{1/2}    
\end{equation}

\begin{equation}\label{eq:18.3}
\epsilon_{1q} \equiv  \left[ \sum_{k=1}^{N} \sum_{j=1}^{N} \epsilon_k \epsilon_j \sin 2 \theta_k \sin 2 \theta_j \cos( \phi_k -  \phi_j)  \right]^{1/2}
\end{equation}

\begin{equation}\label{eq:18.4}
\epsilon_{d} \equiv  \left[ \sum_{k=1}^{N} \sum_{j=1}^{N} \epsilon_k \epsilon_j \sin \theta_k \sin \theta_j \cos (\phi_k - \phi_j)   \right]^{1/2}
\end{equation}

and this allows us to rewrite Eqs. (\ref{eq:17.1}) and (\ref{eq:18.1} in the form

\begin{equation}
P^{(T)} =  \frac{32}{5} \frac{G(1-\zeta)}{c^5} M^2 a^4  \omega^6  \epsilon^2_{2q}   
  + \frac{1}{10} \frac{G(1-\zeta)}{c^5} M^2 a^4 \omega^6  \epsilon^2_{1q} 
 \label{eq:17.1a}
\end{equation}

and
 \begin{equation} \label{eq:18.1a}
 P^{(S)} = \frac{1}{3} \frac{G \zeta}{c^3}  M^2 a^2  \omega^4 \epsilon^2_{d} .
\end{equation}

The ellipticity $\epsilon_{2q}$ corresponds to the GWs emitted at twice the spin-frequency in the tensor wave whereas $\epsilon_{1q}$ corresponds to the waves emitted in tensor waves at the spin-frequency of the pulsar. The ellipticity $\epsilon_{d}$ is related to the scalar GWs. The subscript ‘q’  implies the fact that tensor waves are dominated by time-varying quadrupole moments whereas subscript ‘d’  implies that scalar waves are dominated by time-varying dipole moments.

Since torque is power divided by the angular velocity, we can obtain the following expressions of the torques ($\tau$) exerted by the tensor and scalar waves: 

\begin{equation}
 \tau^{(T)}  =  \frac{32}{5} \frac{G(1-\zeta)}{c^5} M^2 a^4 \omega^5   \epsilon^2_{2q}  + \frac{1}{10} \frac{G(1-\zeta)}{c^5} M^2 a^4 \omega^5  \epsilon^2_{1q}
 \label{eq:19}   
\end{equation}

and
 \begin{equation} \label{eq:20}
 \tau^{(S)} = \frac{1}{3} \frac{G \zeta}{c^3}  M^2 a^2 \omega^3 \epsilon^2_{d}.
\end{equation}

The magnitudes of the rate of change of angular velocity (i.e., angular acceleration) due to tensor and scalar waves are given by

\begin{equation}
 \dot{\omega}^{(T)}  = \frac{32}{5} \frac{G(1-\zeta)}{c^5} \frac{M^2 a^4 \omega^5}{I}  \epsilon^2_{2q}  + \frac{1}{10} \frac{G(1-\zeta)}{c^5} \frac{M^2 a^4 \omega^5}{I} M^2 \epsilon^2_{1q}
 \label{eq:21}   
\end{equation}

and
 \begin{equation} \label{eq:22}
 \dot{\omega}^{(S)} = \frac{1}{3} \frac{G \zeta}{c^3} \frac{ M^2 a^2 \omega^3}{I}  \epsilon^2_{d}.
\end{equation} 
In the above equations, $I$ denotes the moment of inertia of the pulsar about the spinning axis. Note that the spin-down rate of an isolated pulsar can be measured. However, other mechanisms, such as electromagnetic radiation and pulsar wind, also contribute to this spin-down. Therefore, the observed spin-down rate gives an upper limit to the rate due to GW emission.



\section{\label{sec:GR} Results for General relativity}

The general theory of relativity is the limiting case of BD theory, when $\zeta \rightarrow 0$. 
Thus, we can rewrite some of the above results in the case of GR. 
There are only two tensor polarizations in GR, as $h_s (t) = 0$. The strains for `plus' and `cross' polarizations are given by

\begin{eqnarray}    
h^{GR}_+(t)  && = - \frac{2G }{rc^4} \omega^2 M a^2  \left( 1 + \cos^2 \iota \right) \sum_{k=1}^{N} \epsilon_k  \cos \left( 2 \omega t^\prime + 2 \phi_k  \right)   \sin^2 \theta_k   \\
 && + \frac{G }{2rc^4} \omega^2 M a^2 \sin 2 \iota  \sum_{k=1}^{N} \epsilon_k  \cos \left(  \omega t^\prime +  \phi_k  \right) \sin 2 \theta_k 
 \label{eq:11-GR}
\end{eqnarray}
and
\begin{eqnarray}
h^{GR}_{\times}(t) && = -\frac{4G }{rc^4} \omega^2 M a^2 \cos \iota  \sum_{k=1}^{N} \epsilon_k \sin \left( 2 \omega t^\prime + 2 \phi_k  \right) \sin^2 \theta_k   \\
 && + \frac{G }{rc^4} \omega^2 M a^2 \sin  \iota \sum_{k=1}^{N} \epsilon_k \sin \left(  \omega t^\prime +  \phi_k  \right)   \sin 2 \theta_k.
 \label{eq:12-GR}
\end{eqnarray}

\noindent
Consequently, considering no scalar polarization for GR, the expressions for the radiated powers can be written as follows.

\begin{equation}
P_{GR}^{(T)} = \frac{32}{5} \frac{G}{c^5} M^2 a^4 \omega^6   \epsilon^2_{2q}   + \frac{1}{10} \frac{G}{c^5} M^2 a^4 \omega^6  \epsilon^2_{1q}
 \label{eq:17-GR}
\end{equation}
and
 \begin{equation} \label{eq:18-GR}
 P_{GR}^{(S)} = 0.
\end{equation} 

\noindent
These expressions allow us to calculate the expressions of the torque and the angular acceleration, as was done in the case of BD theory.

\section{Single Harmonic model of tensor waves}\label{Single_Harmonic}

In the above sections, we have seen that tensor polarizations are emitted at two frequencies: at the spin-frequency of the pulsar and twice the spin-frequency of the pulsar. If we assume that all deformations lie in the equatorial plane, i.e., $2 \theta_k = n \pi$, the spin-frequency terms in the tensor polarizations vanish. This situation is equivalent to saying that the z-axis in the body axis system coincides with that of the space axis system, assuming that the pulsar is spinning about the z-axis. Under this configuration, an angular velocity in the z-direction would produce an angular momentum only in the z-direction. On substituting $2 \theta_k = n \pi$, expressions for strains for different polarizations take the form

\begin{equation} \label{eq:22.1}
h_+(t)  = - \frac{2G (1-\zeta)}{rc^4} \omega^2 M a^2  \left( 1 + \cos^2 \iota \right) \sum_{k=1}^{N} \epsilon_k  \cos \left( 2 \omega t^\prime + 2 \phi_k  \right)     
\end{equation}

\begin{equation} \label{eq:22.2}
h_{\times}(t)  = -\frac{4G (1-\zeta)}{rc^4} \omega^2  M a^2 \cos \iota  \sum_{k=1}^{N} \epsilon_k \sin \left( 2 \omega t^\prime + 2 \phi_k  \right) 
\end{equation}

\begin{equation} \label{eq:22.3}
h_{S} (t) \approx - \frac{2G}{rc^3} \zeta \omega M a \sin \iota \sum_{k=1}^{N} \epsilon_k  \sin \left( \omega t^\prime + \phi_k \right).
\end{equation}

and the expressions for power emitted in tensor as well as scalar waves become

\begin{equation}
    P^{(T)} =  \frac{32}{5} \frac{G(1-\zeta)}{c^5} M^2 a^4 \omega^6  \sum_{k=1}^{N} \sum_{j=1}^{N} \epsilon_k \epsilon_j  \cos(2 \phi_k - 2 \phi_j)   \nonumber 
 \label{eq:22.4}
\end{equation}

and
 \begin{equation} \label{eq:18}
 P^{(S)} = \frac{1}{3} \frac{G \zeta}{c^3}  M^2 a^2 \omega^4 \sum_{k=1}^{N} \sum_{j=1}^{N} \epsilon_k \epsilon_j  \cos (\phi_k - \phi_j).
\end{equation}

Under this assumption, we observe that scalar waves are emitted at the spin-frequency of the pulsar whereas tensor waves are emitted only at twice the spin-frequency of the pulsar. 

\subsection{Spin-down limit}

Here, we consider the spin-down limit, where we assume that all the rotational kinetic energy lost by the star is through gravitational
radiation. 
Since there are other ways by which pulsars lose energy, with the assumed dominant mechanism being electromagnetic radiation, the spin-down limit gives the maximum GW amplitude that the scalar or tensor polarization can achieve. Spin-down limit plays a vital role in  segregating high-value pulsars in the data. 
A pulsar is classified to be a high value pulsar if the upper limit on the amplitude is less than the spin-down limit.  The upper limit on the amplitude is obtained using data analysis techniques \cite{BD-LVK,BD-LVK1}. 

To address the spin-down limit for tensor polarizations, we assume the simple harmonic model of tensor,
which means that the GWs are emitted at only twice the spin frequency of the pulsar. This
is the case when $\theta_k = \frac{\pi}{2}$ or we can say that all the deformations lie on the equatorial plane.
We can rewrite the `plus’ polarization as: 

\begin{equation} \label{eq:22.1}
h_{+} (t) = -  \frac{1 + \cos^2 \iota}{2} \left[  - h_{01}^q \cos(2 \omega t^\prime ) + h_{02}^q \sin (2 \omega t^\prime )  \right],   
\end{equation} 
where $h_{01}^q$ and $h_{02}^q$ are defined as 

 \begin{equation} \label{eq:22.2}
h_{01}^q \equiv \frac{16 \pi^2 G (1 - \zeta)}{c^4} \frac{f_0^2}{r} M a^2 \sum_{k=1}^{N} \epsilon_k \cos (2 \phi_k)
\end{equation} 

and
 \begin{equation} \label{eq:22.3}
h_{02}^q \equiv \frac{16 \pi^2 G (1 - \zeta)}{c^4} \frac{f_0^2}{r} M a^2 \sum_{k=1}^{N} \epsilon_k  \sin (2 \phi_k).
\end{equation} 

\noindent
Using the same definitions, we can also write `cross' polarization as

 \begin{equation} \label{eq:22.4}
h_{\times} (t) = -  \cos \iota \left[h_{01}^q \sin(2 \omega t^\prime ) + h_{02}^q \cos (2 \omega t^\prime )  \right].   
\end{equation} 
On substitution of Eqs. (\ref{eq:22.4}, \ref{eq:22.1})  into Eq. (\ref{eq:15}), we obtain the following expression of power emitted in tensor waves: 

 \begin{equation} \label{eq:22.4.1}
P^{(T)} = \frac{2}{5} \frac{c^3}{G(1-\zeta)} \omega^2 (h_0^q)^2,
\end{equation} 
where,

 \begin{equation} \label{eq:22.5}
 h_0^q \equiv \sqrt{(h_{01}^q)^2 + (h_{02}^q)^2}.
\end{equation}

\noindent
$h_0^q$ is the dimensionless GW amplitude of the tensor waves. 

If we equate Eq. (\ref{eq:22.4.1}) with the loss of kinetic energy of the pulsar, 

 \begin{equation} \label{eq:22.6}
\frac{d}{dt} \left( I \omega^2 \right) = \frac{2}{5} \frac{c^3}{G(1-\zeta)} \omega^2 (h_0^q)^2,
\end{equation} 
we obtain the spin-down limit in case of tensor waves where superscript `q' denotes the fact that tensor waves are dominated by the time-varying quadrupole moment: 

 \begin{equation} \label{eq:22.7}
h_{0,sd}^q = \frac{1}{r}  \sqrt{\frac{5}{2} \frac{G (1- \zeta)}{c^3} I \frac{|\dot{f}_0|}{f_0}}. 
\end{equation} 

\noindent
Similarly, we can find the spin-down limit for the scalar wave. Let us rewrite the scalar polarization as

 \begin{equation} \label{eq:22.8}
h_s (t) = - \left[h_{01}^d sin (\omega t^\prime) + h_{02}^d cos (\omega t^\prime) \right] \sin \iota ,
\end{equation}
where $h_{01}^d $ and $h_{01}^d$ are defined as:

 \begin{equation} \label{eq:22.9}
h_{01}^d \equiv \frac{4 \pi G}{c^3} \frac{f_0}{r} \zeta  M a \sum_{k=1}^{N}  \epsilon_k  \cos (\phi_k)
\end{equation}
and
 \begin{equation} \label{eq:22.10}
h_{02}^d \equiv \frac{4 \pi G}{c^3} \frac{f_0}{r} \zeta  Ma \sum_{k=1}^{N}   \epsilon_k  \sin (\phi_k).
\end{equation}

\noindent
On substitution of Eq. (\ref{eq:22.8}) into Eq. (\ref{eq:16}), we obtain the expression of power emitted in tensor waves 

 \begin{equation} \label{eq:22.11}
P^{(S)} = \frac{c^3}{12 G \zeta} r^2 \omega^2 (h_0^d)^2,
\end{equation}
where, 
 \begin{equation} \label{eq:22.12}
h_0^d \equiv  \sqrt{(h_{01}^d)^2 + (h_{02}^d)^2}. 
\end{equation}

\noindent
Here, $h_0^d$ is the dimensionless GW amplitude of the scalar waves.

\noindent
If we equate Eq. (\ref{eq:22.11}) to the loss of kinetic energy of the pulsar, we get

 \begin{equation} \label{eq:22.13}
\frac{d}{dt} \left( I \omega^2 \right) = \frac{c^3}{12 G \zeta} r^2 \omega^2 (h_0^d)^2
\end{equation} 
and we obtain the spin-down limit amplitude as  (superscript `d' indicates that scalar waves are dominated by the time-varying dipole moment)

 \begin{equation} \label{eq:22.14}
h_{0,sd}^{d} \equiv \frac{1}{r} \sqrt{\zeta \frac{12 G}{c^3} I \frac{|\dot{f}_0|}{f_0}}.
\end{equation}

Using Eq.(\ref{eq:22.5}) and Eq. (\ref{eq:22.12}), GW strains for the tensorial and scalar waves can be written as

 \begin{equation} \label{eq:22.5.a}
 h_0^q = \frac{16 \pi^2 G (1 - \zeta)}{c^4} \frac{f_0^2}{r} M a^2 \epsilon_{2q}. 
\end{equation}

 \begin{equation} \label{eq:22.12.b}
h_0^d = \frac{4 \pi G \zeta}{c^3} \frac{f_0}{r}   Ma \epsilon_d. 
\end{equation}

\noindent
In the limit of GR, i.e $\zeta \rightarrow 0$, the Eq. (\ref{eq:22.5.a}) takes the form similar to that of Eq. (5) in \cite{BD-LVK}.



\section{Specific examples of mountain distribution}\label{mountain}

In this section, we shall consider two particular example cases, two mountains and four mountains on the pulsar, for the BD theory.

\subsection{Two mountains}\label{Two_mountains}

The net ellipticities, in case of two mountains,  are given by 

\begin{equation} \label{eq:23}
\epsilon_{2q}  =  \left[ \epsilon_1^2 \sin^4 \theta_1 + \epsilon_2^2 \sin^4 \theta_2 + 2 \epsilon_1 \epsilon_2  \sin^2 \theta_1 \sin^2 \theta_2 \cos 2(\phi_1 - \phi_2) \right]^{1/2}  
\end{equation}

\begin{equation} \label{eq:23.1}
\epsilon_{1q}  =  \left[  \epsilon_1^2 \sin^2 2 \theta_1 + \epsilon_2^2 \sin^2 2\theta_2 + 2 \epsilon_1 \epsilon_2  \sin 2\theta_1 \sin 2\theta_2 \cos (\phi_1 - \phi_2)  \right]^{1/2}
\end{equation}

and
\begin{equation} \label{eq:24}
\epsilon_{d} = \left[ \epsilon_1^2 \sin^2 \theta_1 +  \epsilon_2^2 \sin^2 \theta_2 + 2 \epsilon_1 \epsilon_2 \sin \theta_1 \sin \theta_2  \cos (\phi_1 - \phi_2) \right]^{1/2}.
\end{equation}
Let us assume that both mountains lie on the same latitude ($\theta_1 = \theta_2 \equiv \theta$) and that $|\phi_1 - \phi_2| = \frac{\pi}{2} \ \text{or} \  \frac{3 \pi}{2}$. Using Eqs. (\ref{eq:17.1a}) and (\ref{eq:18.1a}), the power emitted in GWs can be written as

\begin{equation} \label{eq:25}
P^{(T)}  = \frac{32}{5} \frac{G(1-\zeta)}{c^5} M^2 a^4 \omega^6 \sin^4 \theta \left[ \epsilon_1 - \epsilon_2  \right]^2  
  + \frac{1}{10} \frac{G(1-\zeta)}{c^5} M^2 a^4 \omega^6 \sin^2 2 \theta \left[ \epsilon_1^2 + \epsilon_2^2\right]
\end{equation}

and
\begin{equation} \label{eq:26}
P^{(S)} = \frac{1}{3} \frac{G \zeta}{c^3} M^2 a^2 \omega^4 \sin^2 \theta \left[ \epsilon_1^2 +  \epsilon_2^2  \right].
\end{equation}
The torques produced by tensor and scalar waves under this configuration are given by

\begin{equation} \label{eq:25.01}
\tau^{(T)}  = \frac{32}{5} \frac{G(1-\zeta)}{c^5} M^2 a^4 \omega^5 \sin^4 \theta \left[ \epsilon_1 - \epsilon_2  \right]^2  
  + \frac{1}{10} \frac{G(1-\zeta)}{c^5} M^2 a^4 \omega^5 \sin^2 2 \theta \left[ \epsilon_1^2 + \epsilon_2^2\right]
\end{equation}

and
\begin{equation} \label{eq:26.01}
\tau^{(S)} = \frac{1}{3} \frac{G \zeta}{c^3} M^2 a^2 \omega^3 \sin^2 \theta \left[ \epsilon_1^2 +  \epsilon_2^2  \right],
\end{equation}
and the corresponding angular accelerations are

\begin{equation} \label{eq:25.02}
\dot{\omega}^{(T)}  = \frac{32}{5} \frac{G(1-\zeta)}{c^5} \frac{M^2 a^4 \omega^5}{I} \sin^4 \theta \left[ \epsilon_1 - \epsilon_2  \right]^2  
  + \frac{1}{10} \frac{G(1-\zeta)}{c^5} \frac{M^2 a^4 \omega^5}{I} \sin^2 2 \theta \left[ \epsilon_1^2 + \epsilon_2^2\right]
\end{equation}

and
\begin{equation} \label{eq:26.01}
\dot{\omega}^{(S)} = \frac{1}{3} \frac{G \zeta}{c^3} \frac{M^2 a^2 \omega^3}{I} \sin^2 \theta \left[ \epsilon_1^2 +  \epsilon_2^2  \right].
\end{equation}
If we assume $\epsilon_1=\epsilon_2 \equiv \epsilon$ and $\theta= \frac{\pi}{2}$, the power emitted in the tensor polarization is zero, and only a scalar wave will carry away the energy from the system, as evident from Eqs. (\ref{eq:25}, \ref{eq:26}).
Similarly, if $|\phi_1 - \phi_2| = \pi$, we obtain

\begin{equation} \label{eq:27}
P^{(T)}  = \frac{32}{5} \frac{G(1-\zeta)}{c^5} M^2 a^4 \omega^6 \sin^4 \theta \left[ \epsilon_1 + \epsilon_2  \right]^2  
  + \frac{1}{10} \frac{G(1-\zeta)}{c^5} M^2 a^4 \omega^6 \sin^2 2 \theta \left[ \epsilon_1 - \epsilon_2\right]^2
\end{equation}

and
\begin{equation} \label{eq:28}
P^{(S)} = \frac{1}{3} \frac{G \zeta}{c^3} M^2 a^2 \omega^4 \sin^2 \theta \left[ \epsilon_1 -  \epsilon_2  \right]^2.
\end{equation}
Here, if $\epsilon_1=\epsilon_2 \equiv \epsilon$, there are no waves at the pulsar spin-frequency from the system and the energy is carried away only by the tensor waves at twice the spin-frequency when both the mountains are at the same latitude as evident from Eqs. (\ref{eq:27}, \ref{eq:28}). 

We can use Eq. (\ref{eq:22.5.a}) to get the GW strain in the presence of two mountains. The  $\epsilon_{2q}$ in this case is given by 

\begin{equation} \label{eq:28.11}
\epsilon_{2q} = \left[ \epsilon_1^2 + \epsilon_2^2  + 2 \epsilon_1 \epsilon_2 \cos (2 \phi_1 - 2 \phi_2)  \right]^{\frac{1}{2}},
\end{equation}

The GW strain, in the case of scalar wave, can be obtained using Eq. (\ref{eq:22.12.b})  where $\epsilon_d$ is given by

\begin{equation} \label{eq:28.21}
\epsilon_d = \left[ \epsilon_1^2 + \epsilon_2^2  + 2 \epsilon_1 \epsilon_2 \cos ( \phi_1 - \phi_2)  \right]^{\frac{1}{2}},
\end{equation}

\noindent
Let us now present the numerical values of the pulsar spin-down rates and strains in Table~\ref{table_1} for the 
following parameter values: $M = 2.8 \times 10^{30}$~kg, $a = 10$~km, $f_0 = 100$~Hz, $r = 4.9 \times 10^{18}$~m. For simplicity, we use $ \zeta = 0.0000125 $ and $I \approx \frac{2}{5} M a^2$. We consider the following cases: 

\begin{itemize}
    \item Case I: $ \epsilon_1 = 1.001 \times 10^{-6}$, $\epsilon_2 = 0$, $\theta_1 = \theta_2 = \frac{\pi}{2}$ and $|\phi_1 - \phi_2| = \frac{\pi}{2}$
    \item Case II: $ \epsilon_1 = 1.001 \times 10^{-6}$,  $ \epsilon_2 = 1.0 \times 10^{-6}$, $\theta_1 = \theta_2 = \frac{\pi}{2}$ and $|\phi_1 - \phi_2| = \frac{\pi}{2}$
    \item Case III: $ \epsilon_1 = \epsilon_2 = 1.0 \times 10^{-6}$, $\theta_1 = \theta_2 = \frac{\pi}{2}$ and $|\phi_1 - \phi_2| = \frac{\pi}{2}$
    \item Case IV:  $ \epsilon_1 = \epsilon_2 = 1.0 \times 10^{-6}$, $\theta_1 = \theta_2 = \frac{\pi}{2}$ and $|\phi_1 - \phi_2| = \pi$
    
\end{itemize}

\begin{table}[h]
\centering
\caption{Examples of numerical values of spin-down rates and strains for parameter values mentioned in section~\ref{Two_mountains}.}
\label{table_1}
\begin{tabular}{|c|c|c|c|c|c|c|}
\hline 
  Case & $\dot{\omega}^{(T)}$ rad $\cdot$ s$^{-2}$ & $\dot{\omega}^{(S)}$ rad $\cdot$ s$^{-2}$ & $ h_0^q $ & $ h_0^d $ &  $e_{2q}$ & $e_{d}$ \\
 \hline
   I  & $1.21 \times 10^{-11}$ & $1.79 \times 10^{-14}$ & $7.44 \times 10^{-25}$ & $2.22 \times 10^{-28}$ & $10^{-6}$ & $10^{-6}$\\
 \hline
   II & $1.20 \times 10^{-17}$ & $3.58 \times 10^{-14}$ & $7.43 \times 10^{-28}$ & $3.14 \times 10^{-28}$ & $10^{-9}$ & $1.41 \times 10^{-6}$ \\
 \hline
    III & $0 $ & $3.57 \times 10^{-14}$ & $ 0 $ & $3.14 \times 10^{-28}$ & $0$ & $1.41 \times 10^{-6}$ \\
 \hline
     IV & $4.82 \times 10^{-11} $ & $0$ & $ 1.49 \times 10^{-24} $ & $ 0 $ & $2 \times 10^{-6}$ & $0$ \\
 \hline
\end{tabular}
\end{table}

\noindent
We emphasize that the $h_0^q$ in Table~\ref{table_1} has been calculated assuming the single harmonic model, which means that tensor waves are emitted only at twice the spin-frequency of the pulsar and $\epsilon_{1q} = 0$ in the single harmonic model. Also, case II suggests how effective ellipticity in the tensor waves reduces to $10^{-9}$ despite having mountains of order $10^{-6}$, as mentioned in the introduction of this paper. 
The numerical values of Table~\ref{table_1} give an idea about the detectability of the GW, as well as the GW contribution to the pulsar spin-down rate. Note that the total spin-down rate could be measured by electromagnetic observations.


Note that $h_0$ and $\dot\omega$ are measured by the detectors. We can also use the strains to estimate the value of the BD parameter $\zeta$. 
Using Eqs.~(\ref{eq:22.5.a}) and (\ref{eq:22.12.b}), we obtain

\begin{equation} \label{eq:val-zeta}
\zeta = \frac{1}{1 + \frac{c}{4 \pi a f_0} \frac{h_0^q \epsilon^d}{h_0^d \epsilon^q}}.  
\end{equation}
In case of only one mountain, $\epsilon^q = \epsilon^d$ and we get

\begin{equation} \label{eq:val-zeta1}
\zeta = \frac{1}{1 + \frac{c}{4 \pi a f_0} \frac{h_0^q }{h_0^d}}.  
\end{equation}


\subsection{Four mountains} 

In the case of four mountains, the effective ellipticity for the scalar and tensor waves takes the following form:

\begin{equation} \label{eq:29.0}
\begin{split}
\epsilon_{2q} & =  \big[\epsilon_1^2 \sin^4 \theta_1 + \epsilon_2^2 \sin^4 \theta_2 + \epsilon_3^2 \sin^4 \theta_3 + \epsilon_4^2 \sin^4 \theta_4 \\
& +  2 \epsilon_1 \epsilon_2  \sin^2 \theta_1 \sin^2 \theta_2 \cos 2(\phi_1 - \phi_2) + 2 \epsilon_1 \epsilon_3  \sin^2 \theta_1 \sin^2 \theta_3 \cos 2(\phi_1 - \phi_3) \\ 
& +   2 \epsilon_1 \epsilon_4  \sin^2 \theta_1 \sin^2 \theta_4 \cos 2(\phi_1 - \phi_4) + 2 \epsilon_2 \epsilon_3  \sin^2 \theta_2 \sin^2 \theta_3 \cos 2(\phi_2 - \phi_3) \\
& +   2 \epsilon_2 \epsilon_4  \sin^2 \theta_2 \sin^2 \theta_4 \cos 2(\phi_2 - \phi_4) + 2 \epsilon_3 \epsilon_4  \sin^2 \theta_3 \sin^2 \theta_4 \cos 2(\phi_3 - \phi_4)
\big]^{1/2}  
\end{split}
\end{equation}

\begin{equation} \label{eq:29.1}
\begin{split}
\epsilon_{1q} & = \big[ \epsilon_1^2 \sin^2 2 \theta_1 + \epsilon_2^2 \sin^2 2\theta_2 + \epsilon_3^2 \sin^2 2\theta_3 + \epsilon_4^2 \sin^2 2\theta_4 \\
 & + 2 \epsilon_1 \epsilon_2  \sin 2\theta_1 \sin 2\theta_2 \cos (\phi_1 - \phi_2) + 2 \epsilon_1 \epsilon_3  \sin 2\theta_1 \sin 2\theta_3 \cos (\phi_1 - \phi_3) \\
 &+ 2 \epsilon_1 \epsilon_4  \sin 2\theta_1 \sin 2\theta_4 \cos (\phi_1 - \phi_4) + 2 \epsilon_2 \epsilon_3  \sin 2\theta_2 \sin 2\theta_3 \cos (\phi_2 - \phi_3) \\
& + 2 \epsilon_2 \epsilon_4  \sin 2\theta_2 \sin 2\theta_4 \cos (\phi_2 - \phi_4) + 2 \epsilon_3 \epsilon_4  \sin 2\theta_3 \sin 2\theta_4 \cos (\phi_3 - \phi_4) \big]^{1/2}
\end{split}
\end{equation}

\begin{equation} \label{eq:29.2}
\begin{split}
\epsilon_{d} & =  \big[ \epsilon_1^2 \sin^2 \theta_1 +  \epsilon_2^2 \sin^2 \theta_2 + \epsilon_3^2 \sin^2 \theta_3 + \epsilon_4^2 \sin^2 \theta_4 \\
& + 2 \epsilon_1 \epsilon_2 \sin \theta_1 \sin \theta_2  \cos (\phi_1 - \phi_2) + 2 \epsilon_1 \epsilon_3 \sin \theta_1 \sin \theta_3  \cos (\phi_1 - \phi_3) \\
& +  2 \epsilon_1 \epsilon_4 \sin \theta_1 \sin \theta_4  \cos (\phi_1 - \phi_4) + 2 \epsilon_2 \epsilon_3 \sin \theta_2 \sin \theta_3  \cos (\phi_2 - \phi_3) \\
& +  2 \epsilon_2 \epsilon_4 \sin \theta_2 \sin \theta_4  \cos (\phi_2 - \phi_4) + 2 \epsilon_3 \epsilon_4 \sin \theta_3 \sin \theta_4  \cos (\phi_3 - \phi_4) 
\big]^{1/2}.
\end{split}
\end{equation}

Now, let us assume that all mountains lie at the same latitude ($\theta_1 = \theta_2 = \theta_3 = \theta_4 \equiv \theta$) such that $| \phi_1 - \phi_2 | =$
$| \phi_1 - \phi_4 | =$ 
$| \phi_2 - \phi_3 | =$
$| \phi_3 - \phi_4 | = \frac{\pi}{2}$,
$| \phi_1 - \phi_3 | = \pi$ and $| \phi_2 - \phi_4 | = \pi$. 
Using Eqs. (\ref{eq:17.1a}) and (\ref{eq:18.1a}) under this assumption, the power emitted in GWs can be written as

\begin{equation} \label{eq:31}
\begin{split}
P^{(T)} & = \frac{32}{5} \frac{G(1-\zeta)}{c^5} M^2 a^4 \omega^6  \sin^4 \theta \left[ \epsilon_1 - \epsilon_2 + \epsilon_3 - \epsilon_4\right]^2  \\
 & + \frac{1}{10} \frac{G(1-\zeta)}{c^5} M^2 a^4 \omega^6  \sin^2 2 \theta \left[ (\epsilon_1 - \epsilon_3)^2 + (\epsilon_2 - \epsilon_4)^2   \right]
\end{split}
\end{equation}

and
\begin{equation} \label{eq:32}
P^{(S)} = \frac{1}{3} \frac{G \zeta}{c^3} M^2 a^2 \omega^4  \sin^2 \theta \left[ (\epsilon_1 - \epsilon_3)^2 + (\epsilon_2 - \epsilon_4)^2   \right].
\end{equation}

The torques produced by tensor and scalar waves under this configuration are given by

\begin{equation} \label{eq:32.01}
\begin{split}
 \tau^{(T)} & = \frac{32}{5} \frac{G(1-\zeta)}{c^5} M^2 a^4 \omega^5  \sin^4 \theta \left[ \epsilon_1 - \epsilon_2 + \epsilon_3 - \epsilon_4\right]^2 \\
 & + \frac{1}{10} \frac{G(1-\zeta)}{c^5} M^2 a^4 \omega^5 \sin^2 2 \theta \left[ (\epsilon_1 - \epsilon_3)^2 + (\epsilon_2 - \epsilon_4)^2   \right]   
\end{split}
\end{equation}

and
\begin{equation} \label{eq:32.02}
\tau^{(S)} = \frac{1}{3} \frac{G \zeta}{c^3} M^2 a^2 \omega^3  \sin^2 \theta \left[ (\epsilon_1 - \epsilon_3)^2 + (\epsilon_2 - \epsilon_4)^2   \right],
\end{equation}
and the corresponding angular accelerations are

\begin{equation} \label{eq:32.03}
\begin{split}
    \dot{\omega}^{(T)}  & = \frac{32}{5} \frac{G(1-\zeta)}{c^5} \frac{M^2 a^4 \omega^5 }{I} \sin^4 \theta \left[ \epsilon_1 - \epsilon_2 + \epsilon_3 - \epsilon_4\right]^2 \\
 & + \frac{1}{10} \frac{G(1-\zeta)}{c^5} \frac{M^2 a^4 \omega^5 }{I} \sin^2 2 \theta \left[ (\epsilon_1 - \epsilon_3)^2 + (\epsilon_2 - \epsilon_4)^2   \right]
  \end{split}
\end{equation}

and
\begin{equation} \label{eq:32.04}
\dot{\omega}^{(S)} = \frac{1}{3} \frac{G \zeta}{c^3} \frac{M^2 a^2 \omega^3 }{I} \sin^2 \theta \left[ (\epsilon_1 - \epsilon_3)^2 + (\epsilon_2 - \epsilon_4)^2   \right].
\end{equation}

It is clear from the above equations that mountains of equal ellipticity will not contribute to any tensor or scalar wave if they are arranged in this configuration for any value of $\theta$.
This shows, as mentioned in section~\ref{sec:intro}, how the effects of multiple mountains can balance each other and reduce the effective ellipticity $\epsilon$  and hence the amplitude of GW strain.
Thus, this example demonstrates the importance of the study of multiple mountains on a pulsar.


  \section{Other realistic mountain distributions}\label{numerical}

In section~\ref{mountain}, we presented toy models to demonstrate how gravitational radiation can be suppressed under some configurations despite some irregularities. In this section, we explore the astrophysical scenario when multiple mountains are present on the surface of the pulsar with different distributions. For all cases/figures, we assume $M = 2.8 \times 10^{30}$~kg, $a = 10$~km, $f_0 = 100$~Hz, $r = 4.9 \times 10^{18}$~m. For simplicity, we use $ \zeta = 0.0000125 $ and $I \approx \frac{2}{5} M a^2$.

\begin{figure}[H]   
\centerline{ \includegraphics[width=15cm]{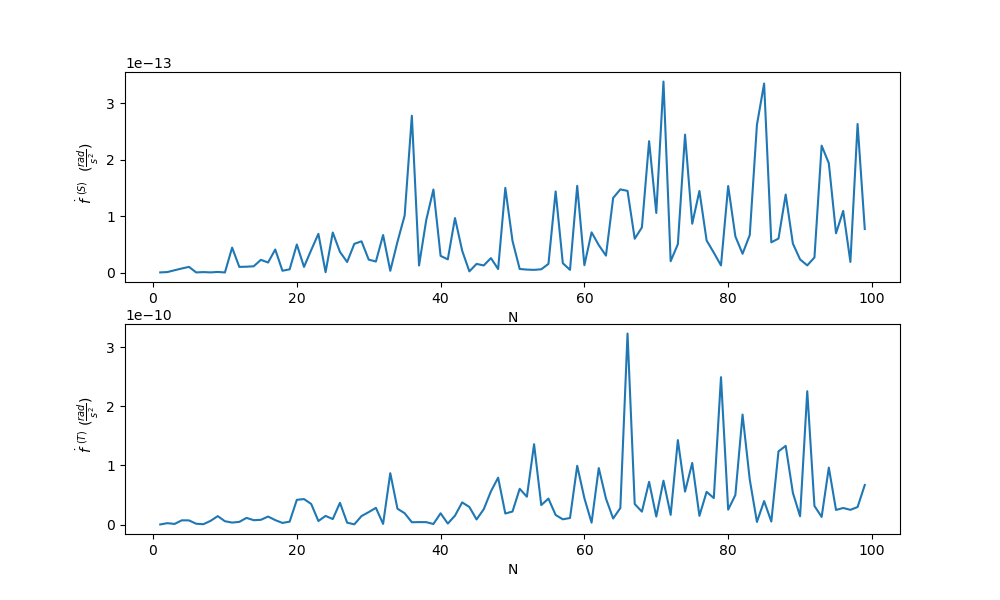} }
\caption{Spin down rate due to scalar waves (top panel) and tensor waves (bottom panel) as functions of number ($N$) of mountains on the surface of a pulsar. 
The locations of mountains are randomly chosen, and mountains have the same ellipticity of $10^{-6}$. }
\label{Fig:1}
\end{figure}

\begin{figure}[H]   
\centerline{ \includegraphics[width=15cm]{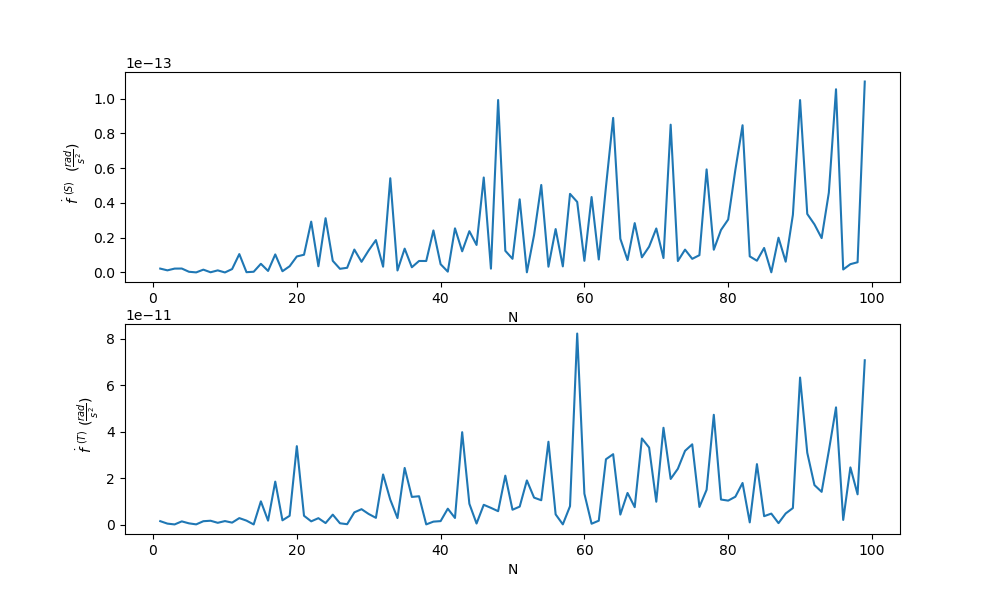} }
\caption{Similar to Figure ~\ref{Fig:1}, but the ellipticity of each mountain is randomly chosen between $0$ and $10^{-6}$.}
\label{Fig:2}
\end{figure}

Figure ~\ref{Fig:1} presents the result of a simulation when N number of mountains of the same mass are present on the surface of a pulsar. 
This represents 
randomly distributed deformations, which cannot be smoothened
by the neutron star’s gravity.
The value of N varies from 1 to 100. The coordinate points for each mountain are obtained by generating random numbers in the respective domain of $(\theta, \phi)$ in spherical coordinates. 
In Figure ~\ref{Fig:1}, the y-axis represents the rate of change of spin-frequency of the pulsar due to the emission of tensor and scalar waves. We observe that the pulsar spin-down rate and the power emitted sensitively depend on the number of mountains and their locations. 
This implies that two very
similar pulsars having a slight difference in number and distribution of mountains can have widely different 
spin-down rates, and one must consider this realistic scenario when comparing models with observations. 
Figure ~\ref{Fig:1} also shows that the power emitted in tensor waves due to time-varying quadrupole moment is typically much larger than the power emitted in scalar waves due to time-varying dipole moment. 
Table ~\ref{table_2} provides the maximum and minimum values of spin-down rates from Figure ~\ref{Fig:1} and the number of mountains corresponding to these extreme values. 
In Figure ~\ref{Fig:2}, we generalize the case of Figure ~\ref{Fig:1} by choosing the ellipticity of each mountain in the range $0$ and $10^{-6}$.
Note that the mass of $i^{th}$ mountain is considered to be $m_i = \epsilon_i M$. 

\begin{table}[h]
\centering
\caption{Table summarizing extreme data points from Figure~\ref{Fig:1}.}
\label{table_2}
\begin{tabular}{|c|c|c|c|c|}
\hline 
  Spin-down rate & Value & No. of mountains ($N$) \\
 \hline
   $\dot{f}^{(T)}_{max}$  & $ 3.2 \times 10^{-10}$ & $66$  \\
 \hline
   $\dot{f}^{(T)}_{min}$  & $ 1.1 \times 10^{-13}$ & $1$  \\
 \hline
    $\dot{f}^{(S)}_{max}$  & $ 3.4 \times 10^{-13} $ & $71$  \\
 \hline
     $\dot{f}^{(S)}_{min}$  & $ 5.5 \times 10^{-16} $ & $ 10$  \\
 \hline
\end{tabular}
\end{table}

\begin{figure}[H]   
\centerline{ \includegraphics[width=15cm]{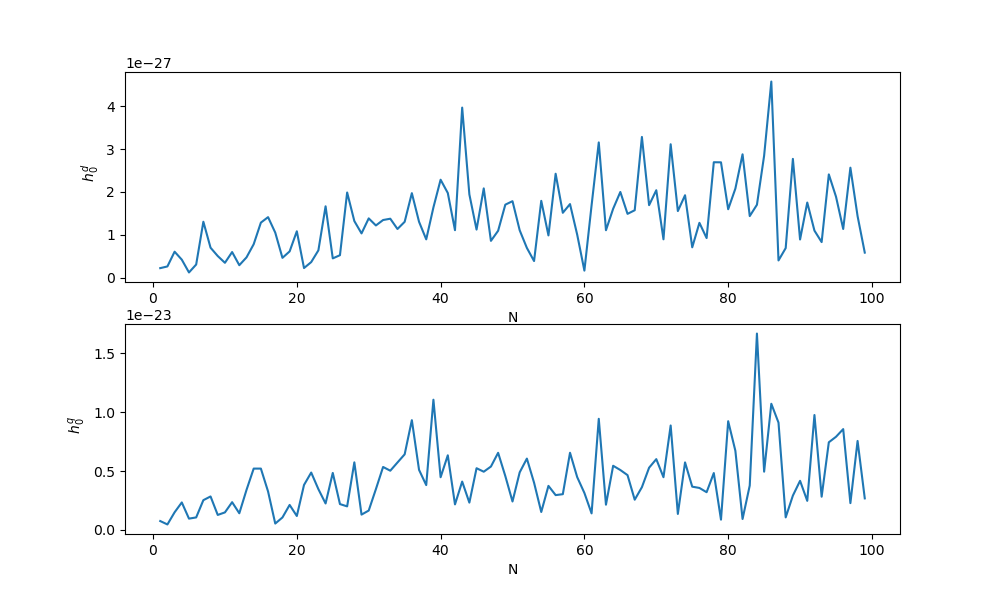} }
\caption{GW amplitude of scalar waves (top panel) and tensor waves (bottom panel) as functions of number ($N$) of mountains on the surface of a pulsar. The locations of mountains are randomly chosen on the equator and mountains have the same ellipticity of $10^{-6}$.}
\label{Fig:h0-1}
\end{figure}

\begin{figure}[H]   
\centerline{ \includegraphics[width=15cm]{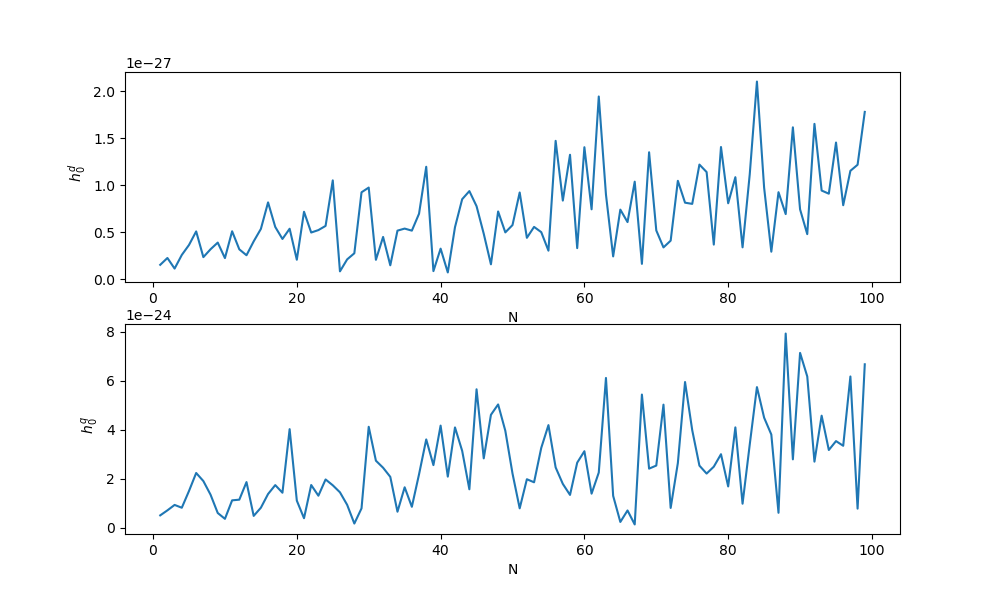} }
\caption{Similar to Figure ~\ref{Fig:h0-1}, but the ellipticity of each mountain is randomly chosen between $0$ and $10^{-6}$.}
\label{Fig:h0-2}
\end{figure}

Figure ~\ref{Fig:h0-1} presents the result of a simulation when N mountains of the same mass are present on the surface of a pulsar. The value of N varies from 1 to 100. The figure corresponds to the single-harmonic model, which means all mountains lie on the equator. The azimuthal angle for each mountain is obtained by generating random numbers in the respective domain of $(\phi)$ in spherical coordinates. In Figure ~\ref{Fig:h0-2}, we generalize the case of Figure ~\ref{Fig:h0-1} by choosing the ellipticity of each mountain in the range $0$ and $10^{-6}$. Table ~\ref{table_3} provides the maximum and minimum values of GW amplitudes from Figure ~\ref{Fig:h0-1} and the number of mountains corresponding to these extreme values.

\begin{table}[h]
\centering
\caption{Table summarizing extreme data points from Figure~\ref{Fig:h0-1}.}
\label{table_3}
\begin{tabular}{|c|c|c|c|c|}
\hline 
  GW amplitude & Value & No. of mountains ($N$) \\
 \hline
   $h^q_{0,max}$  & $ 1.67 \times 10^{-23}$ & $84$  \\
 \hline
   $h^q_{0,min}$  & $ 4.53 \times 10^{-25}$ & $2$  \\
 \hline
    $h^d_{0,max}$  & $ 4.58 \times 10^{-27} $ & $86$  \\
 \hline
     $h^d_{0,min}$  & $ 1.21 \times 10^{-28} $ & $ 5$  \\
 \hline
\end{tabular}
\end{table}

\begin{figure}[H]   
\centerline{ \includegraphics[width=15cm]{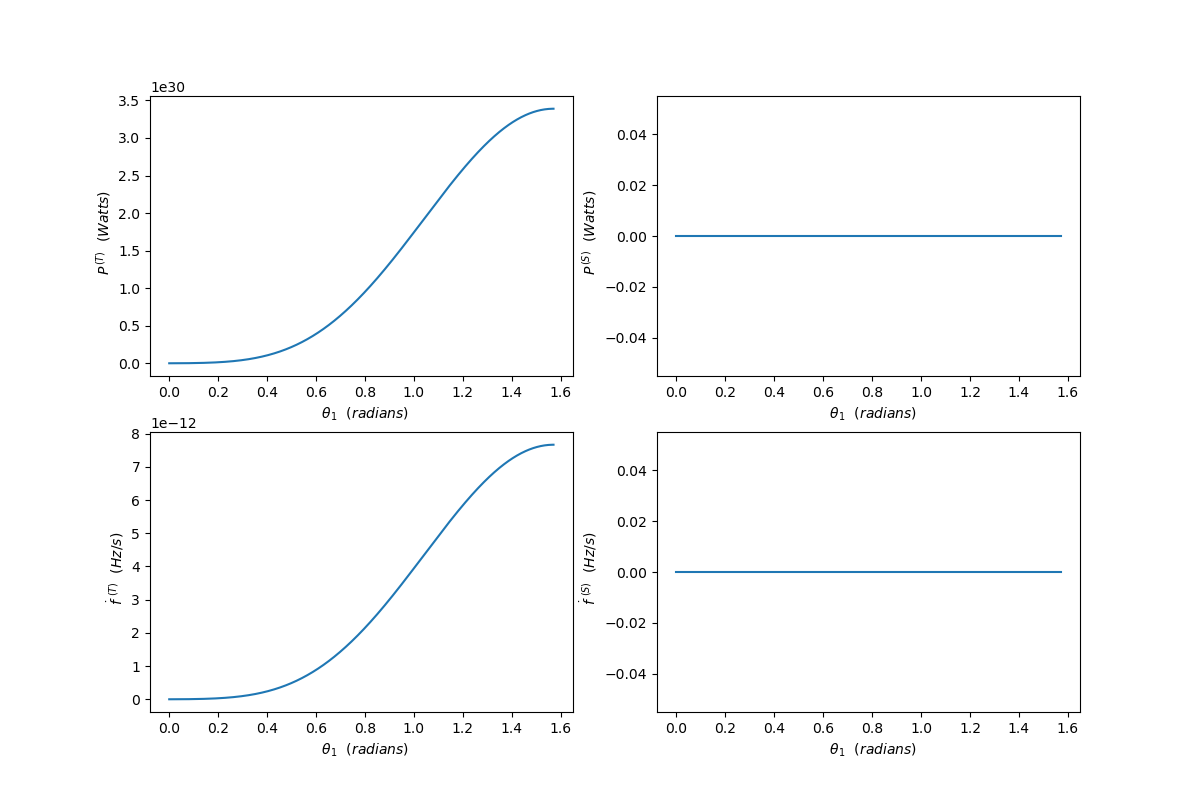} }
\caption{The variation of power emitted and spin-down rate due to two mountains as a function of the latitude ($\theta_1$) of the first mountain. 
The left top panel shows power emitted in transverse waves, the left bottom panel shows spin-down rate due to tensor waves, the right top panel gives power emitted in scalar waves, and the right bottom panel gives spin-down rate due to scalar waves. }
\label{Fig:3}
\end{figure}

Figure ~\ref{Fig:3} presents the results when two mountains are present at antipodal points. 
This case represents mountains (likely due to accretion) at the magnetic poles
of pulsars.
We fix the longitude of the first mountain at $\phi_1 = \frac{\pi}{9}$ and vary the latitude between $0$ and $\frac{\pi}{2}$ in 1000 steps. Corresponding to each step, we obtain the antipodal location for the second mountain and plot the power emitted as well as the spin-down rate as a function of the latitude of the first mountain. We observe that as mountains approach the equator, the power emitted in the tensor wave increases. This, in turn, causes the higher spin-down rate of the pulsar. Interestingly, we observe that there is no power emitted in scalar waves when two mountains lie at the antipodal points. This is because of the fact that the x and y components of the dipole moment of two mountains under this configuration exactly cancel each other (assuming the pulsar is spinning about the z-axis). 

\begin{figure}[H]   
\centerline{ \includegraphics[width=15cm]{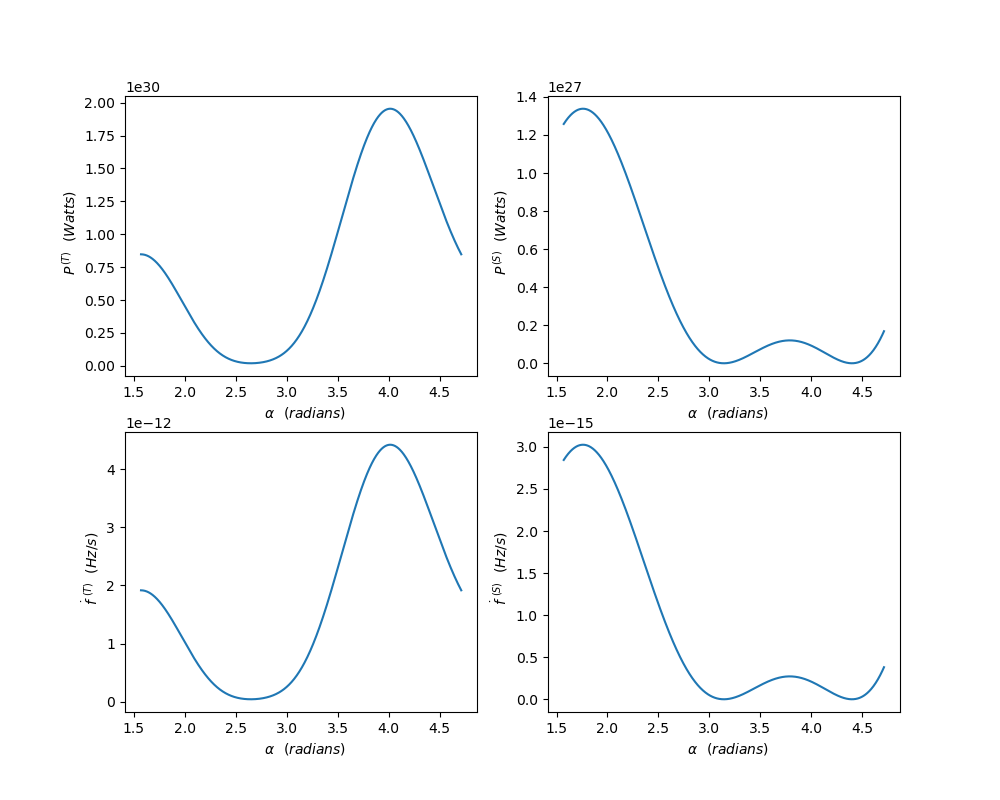} }
\caption{The variation of power emitted and spin-down rate due to two mountains as a function of the difference in their latitudes, $\alpha (\equiv \theta_2 - \theta_1 )$. The value of $\theta_1$ is fixed at $\frac{\pi}{3}$ and $\theta_2$ is obtained by varying $\alpha $ between $\frac{\pi}{2}$ and $\frac{3\pi}{2}$.  The left top panel shows power emitted in transverse waves, the left bottom panel shows spin-down rate due to tensor waves, the right top panel gives power emitted in scalar waves, and the right bottom panel gives spin-down rate due to scalar waves.}
\label{Fig:4}
\end{figure}

Figure ~\ref{Fig:4} presents the results when the difference in latitude ($ \alpha \equiv  \theta_2 - \theta_1$) of two mountains lie in the interval $(\frac{\pi}{2}, \frac{3\pi}{2})$. 
This generalizes the case for Figure ~\ref{Fig:3} and considers that pulsar magnetic poles may not be antipodal \cite{NICER-Riley}.
We fix the value of $\theta_1$ to be $\frac{\pi}{3}$ and vary $\alpha$ between $(\frac{\pi}{2}, \frac{3\pi}{2})$ in 1000 steps. 
The value of $\phi_1$ is set to be $\frac{\pi}{4}$. 
The plot presents power emitted and spin-down rate as a function of $\alpha$. We observe that maximum power in tensor waves is emitted when mountains are approximately 230 degrees apart, whereas minimum power is emitted when mountains are about 152 degrees apart. In the case of scalar waves, maximum power is emitted when mountains are about 101 degrees apart, and minimum power is emitted when the difference is about 252 degrees.

From tables ~\ref{table_1}  and  ~\ref{table_2}, we can easily see that for some specific configurations, the power emitted due to dipole radiation could be higher than that of tensor polarizations. Such configurations could play a vital role in testing theories of gravity with next generation detectors that are sensitive to all six polarization states in a genereic metric theory of gravity.

From Eqs.~(\ref{eq:17.1a}) and ~(\ref{eq:17-GR}), we can compare the power emitted in BD and GR using the relation 

\begin{equation} \label{eq:P-BD-GR}
P^{(T)}_{BD} = \left(1 - \zeta  \right) P^{(T)}_{GR}
\end{equation}
or, 
\begin{equation} \label{eq:P-BD-GR-1}
P^{(T)}_{GR} \approx \left(1 + \zeta  \right) P^{(T)}_{BD}
\end{equation}
Similarly, the relationship between spin-down rates in GR and BD takes the form: 

\begin{equation} \label{eq:om-BD-GR}
\dot{f}^{(T)}_{GR} \approx \left(1 + \zeta  \right) \dot{f}^{(T)}_{BD}
\end{equation}
All figures in this section correspond to BD theory of gravity. 
From Eqs. \ref{eq:P-BD-GR-1} and \ref{eq:om-BD-GR}, one can see that the plots of power emitted or spin-down rate due to tensor polarization in GR can be easily obtained from the case of BD just by multiplying with ($1 + \zeta$). 
However, for GR, $P^{(S)}_{GR} = 0$ and $\dot{f}^{(S)}_{GR} = 0$.

\section{Conclusions}\label{conclusion}

In this paper, we study the effects of multiple mountains on a spinning neutron star or pulsar on gravitational
wave emission and the spin-down rate. 
We consider the Brans–Dicke theory of gravity here, which is more general than general relativity and can reduce to the latter for suitable parameter values.
Typically, the calculations and estimations of continuous gravitational wave emission from a pulsar consider general relativity and an overall ellipticity perpendicular to the spin axis. However, one expects irregularities on the pulsar's surface to the extent the crust can support (see section~\ref{sec:intro}). The effects of these irregularities or mountains could partially balance each other and reduce the gravitational wave emission and the spin-down rate, that is, the overall net ellipticity of the pulsar. This can have overall effects on both electromagnetic and gravitational wave observations.
For example,  the pulsar PSR J1023+0038 cannot have a net ellipticity much greater than $10^{-9}$, even if the observed spin-down rate is entirely due to continuous gravitational waves \cite{SB-2020}. 
But, on the other hand, theoretically the ellipticity can be much higher (see section~\ref{sec:intro}). 
Thus, it is extremely important to consider the realistic scenario of multiple mountains to study the continuous gravitational waves from neutron stars.
In this paper, we introduce this study with astronomically motivated distributions of mountains. 
This will pave the way for more extensive studies in the future.


\acknowledgments
P. V. is partially supported by Grant No. 2022/45/B/ST2/00013 of the National Science Center, Poland.


\appendix

\section{Symmetric Trace Free (STF) tensor} \label{appdx}

In this section, we present the calculation of the symmetric trace-free quadrupole moment of inertia (MOI) tensor in the source frame. 

Let the mass of the $k^{th}$ mountain is $m_k$ and its coordinates are

\begin{eqnarray} 
\label{eq:A1} \nonumber
x_k &=&   a \sin \theta_k \cos \phi_k \\ \nonumber
y_k &=& a \sin \theta_k \sin \phi_k \\ 
z_k &=& a \cos \theta_k
\end{eqnarray} 

The density of the $k^{th}$ mountain is given by

 \begin{equation}\label{eq:A2}
\rho_k = m_k \delta (x-x_k) \delta (y-y_k) \delta (z-z_k)
 \end{equation} 

The $Q^{xx}$ component of the quadrupole tensor can be obtained by

 \begin{equation}\label{eq:A3}
Q^{xx} = Q^{11} = \int \rho \left[ (x^1)^2 - \frac{1}{3} a^2 \right] dx dy dz
 \end{equation} 

 This can be further written as

 \begin{eqnarray}
Q^{xx} && = m \int x^2 \delta(x-x_k) \delta(y-y_k) \delta(z-z_k) \\
&& - \frac{ma^2}{3} \int \delta(x-x_k) \delta(y-y_k) \delta(z-z_k)
\big]
 \label{eq:A4}
\end{eqnarray}

or, 
 \begin{equation}\label{eq:A5}
Q^{xx} = ma^2 \left[ \sin^2 \theta_k \cos^2 \phi_k  - \frac{1}{3}  \right]
 \end{equation} 

 Similarly, other components of the MOI tensor can be obtained when mountains are present. 

 Now we present the calculation of the MOI tensor of a perfect sphere of uniform density $\rho_0$. The component $Q^{xx}$ can be written as
 
  \begin{equation}\label{eq:A6}
Q^{xx} =  \int dm \left[ (x^1)^2 - \frac{1}{3} r^2   \right]
 \end{equation} 

where $ dm $ is the differential mass element given by $ d m = \rho_0 dV$. $dV$ is the differential volume element in spherical coordinates given by $ dV = r^2  \sin \theta d \theta d \phi d r $. Using this, we can write

 \begin{eqnarray}
Q^{xx} && = \rho_0 \int_{r=0}^{a} r^4 dr \int_{\theta = 0}^{\pi} \sin^3 \theta d \theta \int_{\phi = 0}^{2 \pi}  \cos^2 \phi d \phi \\
&& - \frac{1}{3} \rho_0 \int_{r=0}^{a} r^4 dr \int_{\theta = 0}^{\pi} \sin \theta d \theta \int_{\phi = 0}^{2 \pi}  d \phi
 \label{eq:A7}
\end{eqnarray}

or, 

  \begin{equation}\label{eq:A6}
Q^{xx} = 0
 \end{equation}

 Using a similar calculation, we find that all the components of the MOI tensor of a perfect sphere with uniform density vanish. 





\begin{thebibliography}{99}

%
\def\cmp#1#2#3#4{\emph{#4}, \emph{ Commun. Math. Phys.} {\bf #1} (#3) #2}
\def\lmp#1#2#3#4{\emph{#4}, \emph{ Lett. Math. Phys.} {\bf #1} (#3) #2}
\def\hpa#1#2#3#4{\emph{#4}, \emph{ Hell. Phys. Acta} {\bf #1} (#3) #2}
\def\grg#1#2#3#4{\emph{#4}, \emph{ Gen. Rel. Grav.} {\bf #1} (#3) #2}
\def\pr#1#2#3#4{\emph{#4}, \emph{ Phys. Rev.} {\bf #1} (#3) #2}
\def\prl#1#2#3#4{\emph{#4}, \emph{ Phys. Rev. Lett.} {\bf #1}, #2 (#3)}
\def\prd#1#2#3#4{\emph{#4}, \emph{ Phys. Rev. D} {\bf #1}, #2 (#3)}

\def\prb#1#2#3#4{\emph{#4}, \emph{ Phys. Rev. B} {\bf #1}, #2 (#3) }
\def\prx#1#2#3#4{\emph{#4}, \emph{ Phys. Rev. X} {\bf #1} (#3) #2}
\def\pl#1#2#3#4{\emph{#4}, \emph{ Phys. Lett.} {\bf #1} (#3) #2}
\def\pla#1#2#3#4{\emph{#4}, \emph{ Phys. Lett. A} {\bf #1} (#3) #2 }
\def\plb#1#2#3#4{\emph{#4}, \emph{ Phys. Lett. B} {\bf #1}, #2 (#3)}
\def\prep#1#2#3#4{\emph{#4}, \emph{ Phys. Reports} {\bf #1}, #2 (#3)}
\def\phys#1#2#3#4{\emph{#4}, \emph{ Physica} {\bf #1} (#3) #2}
\def\jcp#1#2#3#4{\emph{#4}, \emph{ J. Comput. Phys.} {\bf #1} (#3) #2}
\def\jmp#1#2#3#4{\emph{#4}, \emph{ J. Math. Phys.} {\bf #1} (#3) #2}
\def\jpm#1#2#3#4{\emph{#4}, \emph{ J. Phys. A: Math. Gen.} {\bf #1} (#3) #2}
\def\cpr#1#2#3#4{\emph{#4}, \emph{ Computer Phys. Rept.} {\bf #1} (#3) #2}
\def\cqg#1#2#3#4{\emph{#4}, \emph{ Class. Quant. Grav.} {\bf #1} (#3) #2}
\def\cma#1#2#3#4{\emph{#4}, \emph{ Computers Math. Applic.} {\bf #1} (#3) #2}
\def\mc#1#2#3#4{\emph{#4}, \emph{ Math. Compt.} {\bf #1} (#3) #2}
\def\apj#1#2#3#4{\emph{#4}, \emph{ Astrophys. J.} {\bf #1} (#3) #2}
\def\apjs#1#2#3#4{\emph{#4}, \emph{ Astrophys. J. Suppl.} {\bf #1} (#3) #2}
\def\apjl#1#2#3#4{\emph{#4}, \emph{ Astrophys. J. Lett.} {\bf #1} (#3) #2}
\def\acta#1#2#3#4{\emph{#4}, \emph{ Acta Astronomica} {\bf #1} (#3) #2}
\def\apl#1#2#3#4{\emph{#4}, \emph{ Ann. Physik. (Leipzig)} {\bf #1} (#3) #2}
\def\amjp#1#2#3#4{\emph{#4}, \emph{Am. J. Phys.} {\bf #1} (#3) #2}
\def\anp#1#2#3#4{\emph{#4}, \emph{ Ann. Phys.} {\bf #1} (#3) #2}
\def\sa#1#2#3#4{\emph{#4}, \emph{ Sov. Astro.} {\bf #1} (#3) #2}
\def\sia#1#2#3#4{\emph{#4}, \emph{ SIAM J. Sci. Statist. Comput.} {\bf #1} (#3) #2}
\def\aa#1#2#3#4{\emph{#4}, \emph{ Astron. Astrophys.} {\bf #1} (#3) #2}
\def\mnras#1#2#3#4{\emph{#4}, \emph{ Mon. Not. R. Astr. Soc.} {\bf #1} (#3) #2}
\def\npb#1#2#3#4{\emph{#4}, \emph{ Nucl. Phys. B} {\bf #1}, #2 (#3)}
\def\npa#1#2#3#4{\emph{#4}, \emph{ Nucl. Phys. A} {\bf #1} (#3) #2}

\def\prsla#1#2#3#4{\emph{#4}, \emph{ Proc. R. Soc. London, Ser. A} {\bf #1} (#3) #2}
\def\jhep#1#2#3#4{\emph{#4}, \emph{ JHEP} {\bf #1} (#2) #3}
\def\jcap#1#2#3#4{\emph{#4}, \emph{ JCAP} {\bf #1} (#2) #3}

\def\nuca#1#2#3#4{\emph{#4}, \emph{ Nuovo Cimento A } {\bf #1} (#3) #2}
\def\nucb#1#2#3#4{\emph{#4}, \emph{ Nuovo Cimento B } {\bf #1} (#3) #2}
\def\ijmp#1#2#3#4{\emph{#4}, \emph{ Int. J. Mod. Phys. D} {\bf #1} (#3) #2}
\def\atmp#1#2#3#4{\emph{#4}, \emph{ Adv. Theor. Math. Phys.} {\bf #1} (#3) #2}
\def\ptps#1#2#3#4{\emph{#4}, \emph{ Prog. Theor. Phys. Suppl.} {\bf #1} (#3) #2}
\def\ptp#1#2#3#4{\emph{#4}, \emph{ Prog. Theor. Phys.} {\bf #1} (#3) #2}
\def\lmp#1#2#3#4{\emph{#4}, \emph{ Lett. Math. Phys.} {\bf #1} (#3) #2}
\def\cpam#1#2#3#4{\emph{#4}, \emph{ Comm. Pure Appl. Math.}  {\bf #1} (#3) #2}
\def\adv#1#2#3#4{\emph{#4}, \emph{ Adv. Phys.}  {\bf #1} (#3) #2}
\def\zh#1#2#3#4{\emph{#4}, \emph{ Zh. Eksp. Teor. Fiz.}  {\bf #1} (#3) #2}
\def\mplb#1#2#3#4{\emph{#4}, \emph{ Mod. Phys. Lett. B} {\bf #1}, #2 (#3)}
\def\mpla#1#2#3#4{\emph{#4}, \emph{ Mod. Phys. Lett. A} {\bf #1}, #2 (#3)}


\def\jams#1#2#3#4{\emph{#4}, \emph{ J. Austral. Math. Soc. B} {\bf #1} (#3) #2}
\def\appa#1#2#3#4{\emph{#4}, \emph{ Acta Phys. Polonica A} {\bf #1} (#3) #2}
\def\appb#1#2#3#4{\emph{#4}, \emph{ Acta Phys. Polonica B} {\bf #1} (#3) #2}

\def\nat#1#2#3#4{\emph{#4}, \emph{Nature} {\bf #1} #2 (#3)}
\def\natcom#1#2#3#4{\emph{#4}, \emph{Nature Commun.} {\bf #1} (#3) #2}
\def\natphys#1#2#3#4{\emph{#4}, \emph{Nature Physics} {\bf #1} (#3) #2}
\def\natmat#1#2#3#4{\emph{#4}, \emph{Nature Mat.} {\bf #1} (#3) #2}


\def\science#1#2#3#4{\emph{#4}, \emph{Science} {\bf #1} (#3) #2}
\def\sciadv#1#2#3#4{\emph{#4}, \emph{Sci. Adv.} {\bf #1} (#3) #2}

\def\arcmp#1#2#3#4{\emph{#4}, \emph{Annual Rev. of Cond. Matter Physics} {\bf #1} (#3) #2}
\def\zphys#1#2#3#4{\emph{#4}, \emph{Z. Phys.} {\bf #1}, (#3) #2}
\def\ncs#1#2#3#4{\emph{#4}, \emph{Nuovo Cimento Suppl.} {\bf #1} (#3) #2}
\def\physb#1#2#3#4{\emph{#4}, \emph{Physica B} {\bf #1}, (#3) #2}
\def\jpcm#1#2#3#4{\emph{#4}, \emph{J. Phys.: Condens. Matter } {\bf #1} (#3) #2}
\def\pnas#1#2#3#4{\emph{#4}, \emph{Proc. Nat. Academy Sciences} {\bf #1} (#3) #2}
\def\sssr#1#2#3#4{\emph{#4}, \emph{Izv. Akad Nauk SSSR, ser. fiz.} {\bf #1} (#3) #2}
\def\jpg#1#2#3#4{\emph{#4}, \emph{ J. Phys. G} {\bf #1} (#3) #2}
\def\chinpb#1#2#3#4{\emph{#4}, \emph{Chin. Phys. B} {\bf #1} (#3) #2}
\def\njp#1#2#3#4{\emph{#4}, \emph{ New J. Phys.} {\bf #1} (#3) #2}
\def\frontphys#1#2#3#4{\emph{#4}, \emph{ Front. Phys.} {\bf #1} (#3) #2}
\def\epl#1#2#3#4{\emph{#4}, \emph{ EPL} {\bf #1} (#3) #2}
\def\rmp#1#2#3#4{\emph{#4}, \emph{ Rev. Mod. Phys.} {\bf #1}, #2 (#3)}
\def\rpp#1#2#3#4{\emph{#4}, \emph{ Rep. Prog. Phys.} {\bf #1}, #2 (#3)}

\def\hepph#1#2{{ hep-ph }{#1} (#2)}
\def\arxiv#1#2#3{\emph{#3},{ arXiv }{#1} (#2)}
\def\hepth#1#2{{ hep-th }{#1} (#2)}
\def\grqc#1#2{{ gr-qc }{#1} (#2)}
\def\ibid#1#2#3#4{\emph{#4}, {\it ibid.} {\bf #1} (#3) #2}
\def\conphy#1#2#3#4{\emph{#4}, \emph{Contemporary Physics} {\bf #1}, (#3) #2}
\def\ppnp#1#2#3#4{\emph{#4}, \emph{ Prog. Part. Nucl. Phys} {\bf #1} (#3) #2}
\def\arnps#1#2#3#4{\emph{#4}, \emph{ Annu. Rev. Nucl. Part. Sci.} {\bf #1} (#3) #2}
\def\ijmpa#1#2#3#4{\emph{#4}, \emph{ Int. J. Mod. Phys. A} {\bf #1}, #2 (#3)}
\def\jams#1#2#3#4{\emph{#4}, \emph{ J. Austral. Math. Soc. B} {\bf #1} (#3) #2}
\def\appa#1#2#3#4{\emph{#4}, \emph{ Acta Phys. Polonica A} {\bf #1}, (#3) #2}
\def\nat#1#2#3#4{\emph{#4}, \emph{Nature} {\bf #1}, (#3) #2}
\def\science#1#2#3#4{\emph{#4}, \emph{Science} {\bf #1}, (#3) #2}
\def\arcmp#1#2#3#4{\emph{#4}, \emph{Annual Rev. of Cond. Matter Physics} {\bf #1}, (#3) #2}
\def\jcap#1#2#3#4{\emph{#4}, \emph{JCAP} {\bf #1}, (#3) #2}
\def\conphy#1#2#3#4{\emph{#4}, \emph{Contemporary Physics} {\bf #1}, (#3) #2}
\def\ptps#1#2#3#4{\emph{#4}, \emph{ Prog. Theor. Phys. Suppl.} {\bf #1} (#3) #2}
\def\ptp#1#2#3#4{\emph{#4}, \emph{ Prog. Theor. Phys.} {\bf #1} (#3) #2}
\def\apjsup#1#2#3#4{\emph{#4}, \emph{ Astrophys. J. Suppl. Ser.} {\bf #1} (#3) #2}
\def\eurphysjc#1#2#3#4{\emph{#4}, \emph{ Eur. Phys. J.  C} {\bf #1}, #2 (#3)}
\def\njp#1#2#3#4{\emph{#4}, \emph{ New J. Phys. } {\bf #1} (#3) #2}
\def\eurphysjplus#1#2#3#4{\emph{#4}, \emph{ Eur. Phys. J.  Plus} {\bf #1}, #2 (#3)}
%
\def\hepph#1#2{{ hep-ph }{#1} (#2)}
\def\hepth#1#2{{ hep-th }{#1} (#2)}
\def\astroph#1#2{{ astro-ph }{#1} (#2)}
\def\grqc#1#2{{ gr-qc }{#1} (#2)}
\def\ibid#1#2#3#4{\emph{#4}, {\it ibid.} {\bf #1} (#3) #2}

\def\contp#1#2#3#4{\emph{#4}, \emph{ Contemporary Physics} {\bf #1}, #2 (#3)}
\def\physdarkun#1#2#3#4{\emph{#4}, \emph{ Phys. of Dark Universe } {\bf #1}, #2 (#3)}
\def\astrsc#1#2#3#4{\emph{#4}, \emph{Astrophys. Space Sci.} {\bf #1}, #2 (#3)}

\def\epjc#1#2#3#4{\emph{#4}, \emph{ Eur. Phys. J. C} {\bf #1} #2 (#3) }
\def\revphys#1#2#3#4{\emph{#4}, \emph{Reviews in Phys.} {\bf #1} #2 (#3) }







%





\bibitem{Hask-Bej} B.Haskell and B. M.Bejger. Astrophysics with continuous gravitational waves.  {\em Nature Astronomy} {\bf 2023}, {\em 7}, 1160–1170 

\bibitem{Piccini} O.J Piccinni. Status and Perspectives of Continuous Gravitational Wave Searches.  {\em Galaxies} {\bf 2022}, {\em 10(3)}, 72

\bibitem{JKS1} P. Jaranowski, A. Królak, and B. F. Schutz. Data analysis of gravitational-wave signals from pulsars. I. The signal and its detection.  {\em Phys. Rev. D} {\bf 1998}, {\em 58}, 063001

\bibitem{JK1} P. Jaranowski and A. Królak. Data analysis of gravitational-wave signals from pulsars.II. Accuracy of estimation of parameters.  {\em Phys. Rev. D} {\bf 1999}, {\em 59}, 063003

\bibitem{JK2} P. Jaranowski and A. Królak. Data analysis of gravitational-wave signals from pulsars. III. Detection statistics and computational requirements  {\em Phys. Rev. D} {\bf 2000}, {\em 61}, 062001


\bibitem{Jones2010} Jones, D.I. Gravitational wave emission from rotating superfluid neutron stars. {\em MNRAS} {\bf 2010}, {\em 402}, 2503–2519

\bibitem{astone2012} Astone P., Colla A., D'Antonio S., Frasca S. and Palomba C, Coherent search of continuous gravitational wave signals: extension of the 5-vectors method to a network of detectors, {\em MNRAS} {\bf 2012}, {\em 363}, 012038


\bibitem{pitkin2017} M. Pitkin, M. Isi, J. Veitch and G. Woan, A NESTED SAMPLING CODE FOR TARGETED SEARCHES FOR CONTINUOUS GRAVITATIONAL WAVES
FROM PULSARS, arXiv:1705.08978 [gr-qc]

\bibitem{GR test GWTC3} Abbott, R.; Abe, H.; Acernese, F.; Ackley, K.; Adhikari, N.; Adhikari, R.X.; Adkins, V.K.; Adya, V.B.; Affeldt, C.; Agarwal, D.; et al. Tests of General Relativity with GWTC-3. arXiv:2112.06861


\bibitem{GR-test-GW170817} Abbott, B.P; Abbott, R.; Abbott, T.D; Acernese, F.; Ackley, K.; Adams, C.; Adams, T.; Addesso, P.; Adhikari, R.X.; Adya, V.B.; et al. Tests of General Relativity with GW170817. {\em Phys. Rev. Lett.} {\bf 2019}, {\em 123}, 011102.


\bibitem{Wex} Wex, N. Testing Relativistic Gravity with Radio Pulsars. arXiv:1402.5594 

\bibitem{Will2014} Will, C.M. The Confrontation between General Relativity and Experiment. {\em Living Rev. Relativ.} {\bf 2014}, {\em 17}, 4


\bibitem{Brans-Dicke} Brans, C.; Dicke, R.H.  Mach's Principle and a Relativistic Theory of Gravitation. {\em Phys. Rev. Lett.} {\bf 1961}, {\em 124}, 925



\bibitem{Jordan} Jordan, P. Zum gegenw\"artigen Stand der Diracschen kosmologischen Hypothesen. {\em Z. Phys} {\bf 1959}, {\em 157}, 112–121.

\bibitem{Fierz} Fierz, M.  \"Uber die physikalische Deutung der erweiterten Gravitationstheorie P.
Jordans. {\em Helv. Phys. Acta} {\bf 1956}, {\em 29}, 128–134


\bibitem{Misner}Misner, C.W.; Thorne, K.S.; Wheeler, J.A.  {\em Gravitation}; W. H. Freeman and Company, San Francisco, USA, 1973 

\bibitem{Fuji-Maeda} Fuji, Y.; Maeda K. {\em The scalar-tensor theory of gravitation }, Cambridge University Press, Cambridge, UK, 2004.  



\bibitem{Hongsu} H. Kim, Brans-Dicke theory as a unified model for dark matter-dark energy, {\bf 2005}, {\em 364}, 813–822. 

\bibitem{Bertolami} Bertolami, O.; Martins, P.J. 
Nonminimal coupling and quintessence {\em Phys. Rev. D} {\bf 2000}, {\em 61}, 064007.

\bibitem{JBD-cosmo} Almeida, C.R.; Galkina, O.; Fabris, J.C. Quantum and Classical Cosmology in the Brans–Dicke Theory. {\em Universe}, {\bf 2021}, {\em 7}, 286.

\bibitem{Sola} Sola, J.; Gomez-Valent, A.; Perez, J.D.C.; Moreno-Pulido, C. Brans-Dicke cosmology with a $\Lambda$ - term: a possible solution to $ \Lambda $CDM tensions. {\em Class. Quantum Grav}, {\bf 2020}, {\em 37}, 245003.





\bibitem{Will-1977} Will, C.M. Gravitational radiation from binary systems in alternative metric theories of gravity: dipole radiation and the binary pulsar. {\em ApJ.} {\bf 1977}, {\em 214}, 826-839. 

\bibitem{Verma} Verma, P. Probing Gravitational Waves from Pulsars in Brans-Dicke Theory. {\em Universe} {\bf 2021}, {\em 7(7)}, 235

\bibitem{BD-LVK} Abbott, R.; Abe, H.; Acernese, K.; Ackley, N.; Adhikari, N.; Adhikari, R.X.; Adkins, V.K.; Adya, V.B.; Affeldt, C.; Agarwal, D.; et al. Searches for Gravitational Waves from Known Pulsars at Two Harmonics in the Second and Third LIGO-Virgo Observing Runs. {\em APJ} {\bf 2022}, {\em 935:1}, 29pp


\bibitem{BD-LVK1} Abac, A.G.; Abbott, R.; Abouelfettouh, I.;  Abe, H.; Acernese, K.; Ackley, N.; Adhikari, N.; Adhikari, R.X.; Adkins, V.K.; Adya, V.B.; Affeldt, C.; Agarwal, D.; et al. Search for continuous gravitational waves from known pulsars in the first part of the fourth
LIGO-Virgo-KAGRA observing run. {\em APJ} {\bf 2025}, {\em 983}, 99



\bibitem{SB-2017} Bhattacharyya, S.; Bombaci, I.; Bandyopadhyay, D.; Thampan, A.V.; Logoteta, D.; Millisecond radio pulsars with known masses: Parameter values and equation of state models. {\em New Astronomy} {\bf 2017}, {\em 54}, 61-71.

\bibitem{NICER-Riley} Riley, T.E et. al. A NICER View of PSR J0030+0451: Millisecond Pulsar Parameter Estimation. {\em APJ} {\bf 2019}, {\em 887:L21}, 60pp.

\bibitem{BH-2006} Haskell, B.; Jones, D.I.; Andersson, N. Mountains on neutron stars: accreted versus non-accreted crusts. {\em MNRAS} {\bf 2006}, {\em 373}, 1423–1439.

\bibitem{BH-2017} Haskell, B.; Patruno, A. Are Gravitational Waves Spinning Down PSR J1023 + 0038? {\em PRL} {\bf 2017}, {\em 119}, 161103. 

\bibitem{Woan-2018} Woan, G.; Pitkin, M.D.; Haskell, B.; Jones, D.I.; Lasky, P.D. Evidence for a Minimum Ellipticity in Millisecond Pulsars. 
{\em ApJL} {\bf 2018}, {\em 863}, L40. 


\bibitem{SB-2020} Bhattacharyya, S. The permanent ellipticity of the neutron star in PSR J1023+0038. {\em MNRAS} {\bf 2020}, {\em 498}, 728–736.


\bibitem{Poisson}Poisson, E and Will, C.M. {\em Gravity Newtonian, Post-Newtonian, Relativistic}; Cambridge University Press, Cambridge, UK, 2014 

\bibitem{Isi} Isi, M.; Pitkin, M.; Weinstein, A.J.  Probing Dynamical Gravity with the Polarization of Continuous Gravitational Waves. {\em Phys. Rev. D.} {\bf 2017}, {\em 96}, 042001.

\bibitem{Verma-rod-BD} Verma, P. A swinging rod in Brans-Dicke theory. {\em Annalen Der Physik} {\bf 2022}, 2100600.  


\bibitem{Cassini} Bertotti, B.; Iess, L.; Tortora, P. A test of general relativity using radio links with the Cassini spacecraft. {\em Nature} {\bf 2003}, {\em 425}, 374–376. 

\bibitem{gw-monopole} Kutschera, M. Monopole gravitational waves from relativistic fireballs driving gamma-ray bursts. {\em MNRAS} {\bf 2003}, {\em 345}, L1–L5.




\bibitem{Krolak} Kr\'olak, A; Jaranowski, P. {\em Analysis of Gravitational-Wave Data}, Cambridge University Press, Cambridge, UK, 2009. 










  


















 














\end{thebibliography}
\end{document}